\documentclass[12pt]{iopart}
\usepackage{graphicx}
\usepackage{iopams}  
\begin{document}

\title[5th LISA symposium]
{Understanding the fate of merging supermassive black holes}

\author{Manuela Campanelli  
%\footnote[3]{To whom correspondence should be addressed (manuela@phys.utb.edu)}
}

\address{Center of Gravitational Wave Astronomy
and Department of Physics and Astronomy, The University of Texas at
Brownsville, Brownsville, Texas 78520}

\begin{abstract}

Understanding the fate of merging supermassive black holes in galactic
mergers, and the gravitational wave emission from this process, are
important LISA science goals. To this end, we present results from
numerical relativity simulations of binary black hole mergers using
the so-called Lazarus approach to model gravitational radiation from
these events. In particular, we focus here on some recent calculations
of the final spin and recoil velocity of the remnant hole formed at
the end of a binary black hole merger process, which may constraint
the growth history of massive black holes at the core of galaxies and
globular clusters.

\end{abstract}


%\submitto{\CQG}

%\maketitle

%SMBH observational evidence
%SMBH-SMBH observational evidence
%LISA sources
%Astrophysics, GR tests
%Source modeling
%Lazarus results
%Final spin, astrophysical consequences
%Recoil velocities, astrophysical consequences
%Conclusions

\section{Introduction}

One of the most exciting scientific objectives for LISA will be to
provide a detailed understanding of the coalescence and gravitational
wave emission processes from merging supermassive black holes in
galactic nuclei. The high expected signal-to-noise ratios make these
astrophysical events `visible' to LISA, up to very high
redshifts. Supermassive binary black hole merger events may therefore
represent the key to understanding the formation and merger histories
of galaxies and their massive central holes. If supermassive black
holes in galactic cores are produced primarily by successive mergers,
rather than accretion, then the details of the merger process may also
have an impact on the population statistics, particularly the
distribution of spins.

The first observational evidence of a galaxy with a pair of active
galactic nuclei has recently been provided by X-ray observations of
NGC 6240 with the Chandra Observatory \cite{Komossa:2002tn}.
% Very recently, a second intermediate mass black hole has been 
% discovered at the centre of our own Galaxy \cite{Maillard:2004wi}.
Possible evidence of merger events has also been presented in radio
observations of X-shaped jet morphologies, which may have been
produced by a sudden change in the central black hole's spin axis
caused by a supermassive black hole - black hole merger
\cite{Merritt:2002hc}.

To interpret the observations made by LISA we will require detailed
theoretical models, based on Einstein's general theory of relativity.
Comparisons with accurate numerical simulations will reveal the
masses, spins, and orientations of the two black holes, providing not
only crucial information about the history and formation of the binary
system but also an important precision test of dynamical nonlinear
general relativistic gravity.

In this paper we briefly review the results from numerical relativity
simulations of binary black hole mergers using the `Lazarus' approach
to model the final moments of a binary black hole merger. In
particular, we focus here on some recent results of the final spin of
the black hole remnant formed at the end of a merger process starting
from moderately spinning binary black holes.  The final spin of
merging massive black holes is crucially related to the birth and
growth history of these systems. We finally present preliminary
calculations about recoil velocities for black hole systems of
comparable but not equal masses. While modest recoil velocities are
known to be sufficient to eject most coalescing black holes from dwarf
galaxies and globular clusters, it is not yet completely understood
how these net velocities can influence the growth history of massive
black holes at the center of giant elliptical and spiral galaxies.

\section{The Lazarus approach}

The Lazarus approach is an effort to explore the modeling of strong
field dynamics and radiation generation in the last moments of a
binary black hole merger by implementing a hybrid approach, which
takes advantage of the complementary strengths and weaknesses of
numerical relativity, black hole perturbation theory and eventually
post-Newtonian approaches. In particular, we follow a full numerical
relativity simulation of a binary black hole system merging into a
single distorted black hole and, at some transition time $T$, extract
data from the simulation which is consequently evolved with black hole
perturbation theory in the close limit approximation
\cite{Price:1994pm}. This approach was successfully developed in Refs
\cite{Baker00b,Baker:2001sf,Baker:2001nu,Baker:2002qf,Baker:2004wv}
and provided the first approximate theoretical estimates for the
gravitational radiation waveforms and energy that are generated during
the plunge of binary black hole systems.

For non-spinning binary black holes mergers, starting from near the
innermost stable circular orbit (ISCO) at a proper separation of
$L\sim 8M$, a burst of radiation is generated during the plunge which
lasts for less than half of an orbital period, carrying away about
$3\%$ of the system's energy and $12\%$ of its angular momentum. At
the end of the coalescence process one obtains a Kerr remnant with a
rotation parameter $a/M\approx 0.72$. A significant qualitative result
is the observation that the waveforms have a relatively simple
form. The dominant part of the waveforms show a circular polarization
pattern like that for a rotating body, and with a smoothly increasing
frequency (up to the quasinormal ringing frequency) and smoothly
varying amplitude. These results are consistent with the
results recently found in Ref. \cite{Alcubierre:2004hr}. 
It is also interesting to note that for proper separations slightly 
larger than those studied here and similar binary black hole 
initial data, the black holes are found to be in orbit 
around each other \cite{Bruegmann:2003aw}.

\section{On the final spin of the remnant black hole}

A consequence of the no hair theorem (or uniqueness theorem) is that
astrophysical black holes can be completely characterized by their
mass $M$ and spin parameters $S=aM$, where $0\leq a/M\leq 1$. Spins in
massive black holes can be produced by a diversity of mechanisms: the
collapse of massive gas accumulations, gas accretion and capture of
stellar mass bodies, successive mergers with other massive holes.
Therefore the spin of the final black hole remnant is important to
understand the supermassive black hole growth history.

%Some observations suggests that black holes may spin as rapidly as
%$a/M\sim 0.98$ (nearly maximally). 

Recent models using the combined effects of gas accretion and binary
coalescences suggest that in effect accretion can be in effect very
efficient to spin up the holes to $a/M\geq 0.8$ and that black holes
may be rapidly rotating in all epochs \cite{Volonteri:2004cf}.  In the
merger scenario spins play an important role. Spins likely drive the
outflows of jets in the core of active galaxies \cite{Rees:1984si},
and appear to be randomly oriented with respect to the plane of the
galactic disk
\cite{Kinney:2000,Kinney:2002}. Black hole mergers may realign the
spin of the more massive hole, inducing a spin flip of the jet in
X-shaped radio morphologies \cite{Merritt:2002hc}.

In Ref.\cite{Baker:2004wv} we studied the merger of two individual
equal mass black holes with spins varying only slowly as the system
approaches the ISCO.  We consider initial spins in the range
$-0.3<s/m_H<0.2$ where $s$ and $m_H$ are respectively the spin and the
apparent horizon mass of each individual hole.  The negative and
positive values indicate respectively spins aligned and
counter-aligned with the orbital angular momentum so we will only see
the non-precessional effects of the strong field spin interactions.
We started our exploratory simulations from sequences of black holes
in circular orbits, based on the assumption that the individual black
hole spins are not strongly affected by interactions, and thus remain
constant through the orbital dynamics.  For both types of data,
circular orbits, and an ISCO can be identified by minimizing the
energy of the system at a fixed separation \cite{Pfeiffer:2000um}.
Note that recent calculations find a somewhat less tightly bound ISCO
\cite{Cook:2004kt}. We then applied the Lazarus approach to treat
several configurations in the described range of initial spins,
allowing us to calculate plunge waveforms as well as radiative loss of
energy and angular momentum.

We show here that for spinning holes the resulting waveforms have
still the same qualitative simple appearance of the non-spinning
binaries supporting the idea that a total mass rescaling of the latter
waveforms can produce an approximate description for both the aligned
and counter-aligned spinning binaries cases.

\begin{figure}
\begin{center}
\includegraphics[width=5.0in]{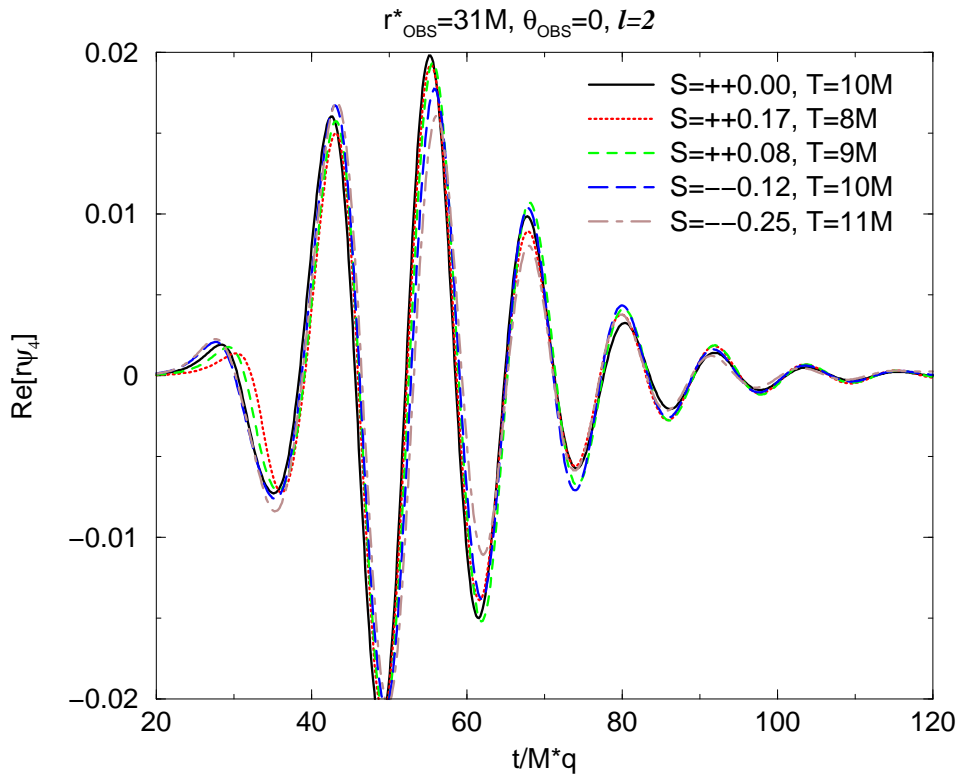}
\end{center}
\caption{Real part of the plunge waveforms from the coalescence of
several spinning black holes for the $l=2$ multipole, as
viewed by an observer located along the polar axis. The waveforms have been
rescaled by a factor $q=\omega_{QN}^{s}/\omega_{QN}^{s=0}$.  $T$
indicates the time after which one can treat each system
perturbatively as described in the Lazarus approach.}
\label{fig:rescala}
\end{figure}

\begin{figure}
\begin{center}
\includegraphics[width=5.0in]{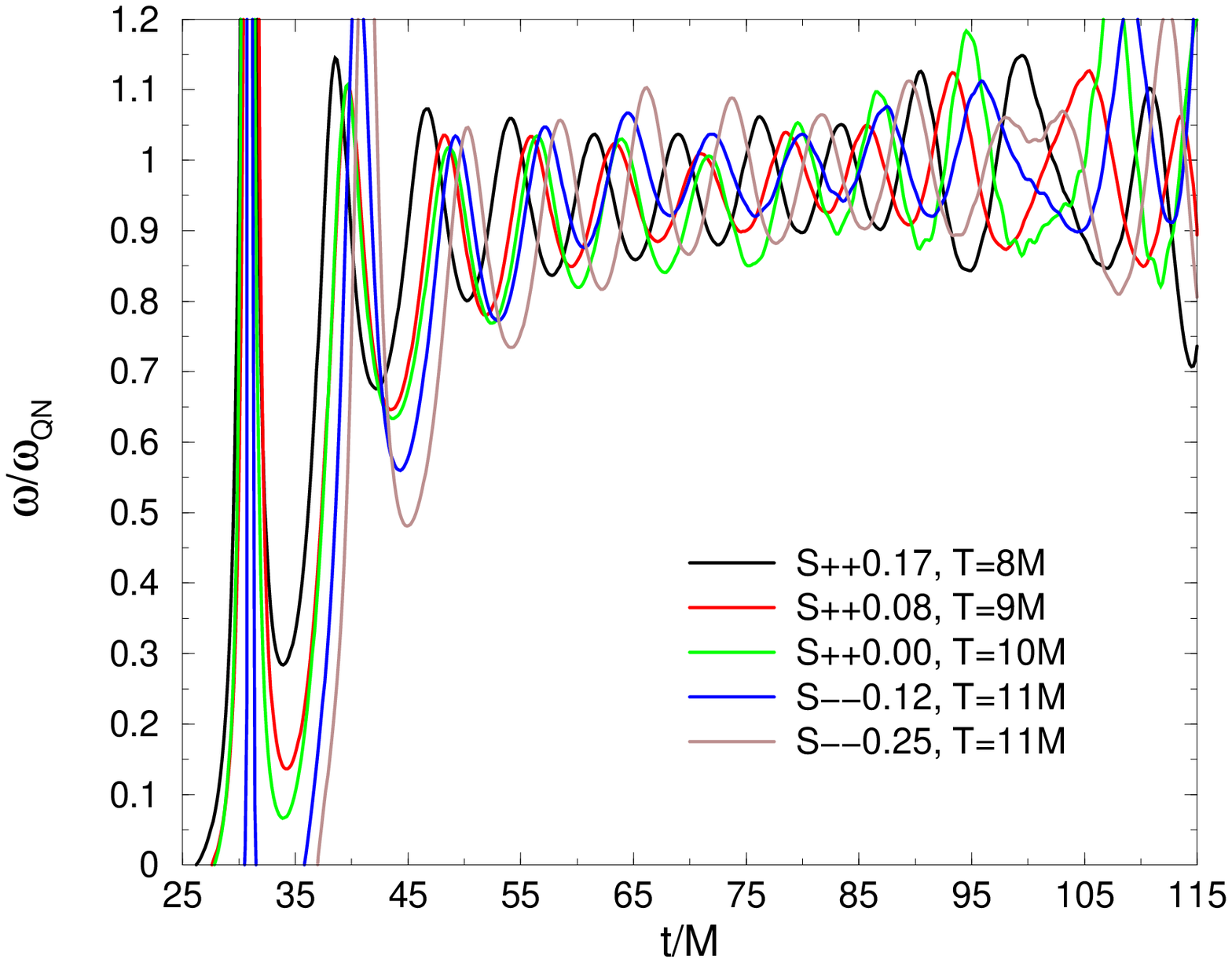}
\end{center}
\caption{
Instantaneous frequency $\omega$ of waveforms described 
in Fig.\ \ref{fig:rescala} normalized to quasinormal frequencies
$\omega_{QN}$ of the remnant Kerr holes. We see that the quasinormal
ringing frequency is approached similarly in each case with a slightly
delayed approach for the counter-aligned cases. This is expected for
these plunge waveforms since the anti-aligned spin initial data starts
with black holes with larger separations than the aligned spin
configurations.
}\label{fig:freq}
\end{figure}

In Fig.\ref{fig:rescala} we superpose the waveforms after a rescaling
of the time axis by $q=\omega_{QN}^{s}/\omega_{QN}^{s=0}$ which
corresponds to rescaling by mass in the $s=0$ case, where
$\omega_{QN}$ is the least damped quasinormal frequency of the final
spinning black hole \cite{Echeverria89}.  We observe a good
superposition, particularly in the ring-down region of the
waveform. The ring-up shows some deviations from the quasinormal
rescaling, not surprisingly, since this part of the waveform is more
affected by the orbital history of the binary system.  This similarity
is consistent with the observed trend in the instantaneous circular
polarization frequency decomposition of the radiation waveforms
displayed in Fig.\ref{fig:freq}.

Concerning the spin of the final Kerr remnant, we find a rotation
parameter $a/M \sim 0.72+0.32 s/m_H$. Crude extrapolations of these
results seem to indicate that nearly maximal rotating black holes with
spins parallel to the orbital angular momentum are needed to form a
maximally spinning remnant hole. It is worth to remark here that our
results are entirely consistent with the approximate results in
Ref.\cite{Gammie:2003qi} indicating that the remnant spin may be as
large as $a/M \sim 0.8-0.9$.  The latter estimates are based on
assumptions that mass and angular momentum are conserved during the
entire coalescence process and using different approximations than our
to find the radius of the ISCO.
 
Complementary studies of binaries in the small mass ratio limit
\cite{Hughes:2002ei} suggest that it is hard to form a maximally
rotating black hole even in the most favorable case of prograde
orbits. The capture of smaller companions with randomly-oriented
individual spins tends to spin the hole down.  See
Ref.\cite{Merritt:2004gc} for a review on the subject.

In conclusion, it is plausible, though not excluded, that binary black
hole mergers are not the most efficient mechanism to produce rapidly
rotating black holes.  Of course rapidly rotating black holes may
still be efficiently produced via gas accretion
\cite{Bardeen:1970,Volonteri:2004cf}.  We are currently working with
more advanced and accurate numerical relativity simulations to
generalize our black hole merger results to nearly equal mass and
rapidly spinning holes.

\section{Recoil velocities}

Gravitational radiation from the coalescence of unequal mass black
hole binaries carries linear momentum, causing the center of mass of
the system to recoil.  There is great astrophysical interest in
determining the gravitational radiation recoil of binary black holes.
One of the questions one can try to answer is whether binary black
holes can be ejected from galaxies. For a galaxy like ours the escape
velocity is of the order of $300 km/s$, while for dwarf galaxies and
globular clusters the escape velocity is a few tens of $km/s$. For
black holes to escape from large galaxies the recoil effects have to
reach $1000 km/s$ \cite{Padmanabhan3}. Redmount and Rees
\cite{1989ComAp..14..165R} discuss some astrophysical implications of
gravitational radiation 'rocket effects'.  Merritt
\cite{Merritt:2002hc} noted that the off centered nuclei observed in
some dwarf elliptical galaxies could be explained by such displacement
of the black central holes. Other possible effects include changes in
the AGN activity, smearing of the central density cusp, distribution
and growth of supermassive black hole masses. It may also determine
the formation rate of intermediate black holes by ejection of a binary
after multiple mergers.

For extreme mass ratios ($q\equiv m_1/m_2\ll 1$) one can use black
hole perturbation theory to describe a test mass inspiraling into a
massive Kerr hole. A crude extrapolation these results to $q\approx
0.4$ suggest that the recoil never exceeds $\sim 500 km/s$
\cite{Merritt:2004xa}. Fitchett and Detweiler
\cite{1984MNRAS.211..933F} have studied a particle in circular orbits
around a Schwarzschild black hole and obtain an estimate of $10km/s$
and an extrapolation to comparable masses leads up to $120 km/s$.  The
post-Newtonian approximation also leads to small recoil
velocities \cite{Wiseman:1992dv}, but does not treat accurately the
plunge phase which is the most relevant in this case.

To study black hole binary systems of comparable masses ($q\approx 1$)
one has to make use of full numerical relativity. Since the recoil
effects vanishes at both, the equal mass and small mass ratio limits,
the maximum linear momentum radiated happens at intermediate mass
ratios. The expected range of $q\approx1/2-1/3$ for the mass ratios
makes numerical relativity to play a crucial role in determining its
precise value. Very little is known for these cases.  In the headon
collision of two comparable mass black holes the radiated linear
momentum was first computed using the close limit approximation
\cite{Andrade:1997pc} and for several larger initial separations using
full 2D nonlinear general relativity in Ref.\cite{Anninos98a}.

\begin{figure}
\begin{center}
\includegraphics[width=5in]{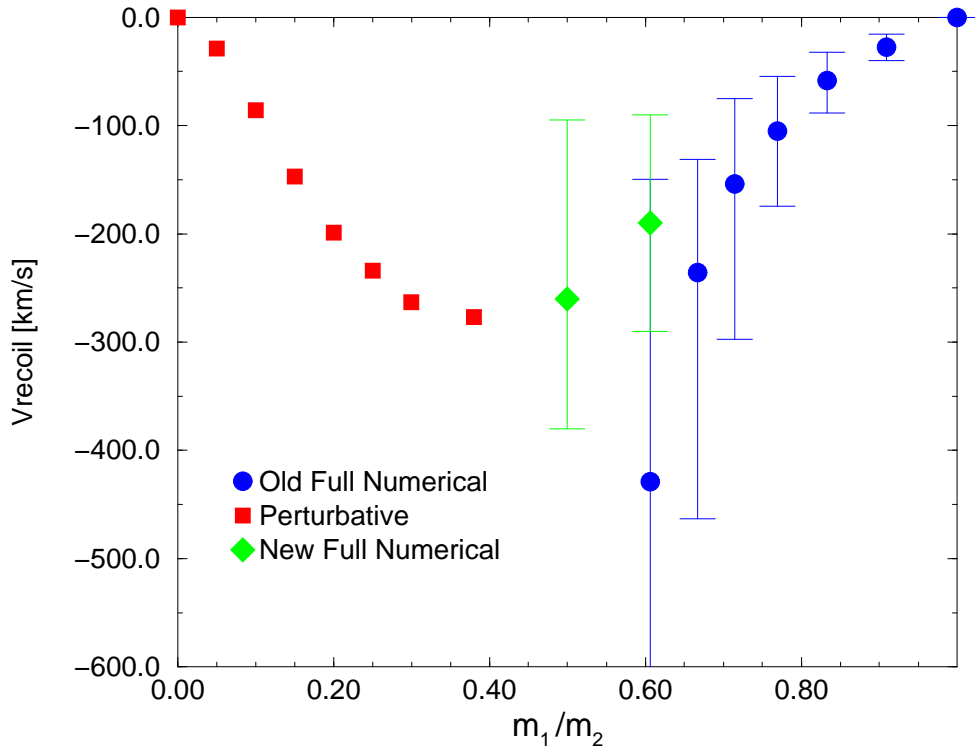}
\end{center}
\caption{Recoil velocity as a function of the mass ratio for
perturbative and full numerical calculations. The perturbative
values are from Ref.\cite{Favata:2004wz}. The error bars 
in the numerical calculations are computed taking the 
maximum and the minimum of the values of the total 
linear momentum as a function of the transition time $T$ 
from non-linear to linear evolution. The minus sign 
convention is because we actually compute 
the momentum carried out by the gravitational waves.}
\label{fig:PxM}
\end{figure}

In this paper we present some preliminary investigations of numerical
simulations of unequal but comparable mass non-spinning black holes
starting from the ISCO as computed in Ref. \cite{Pfeiffer:thesis}. We
evolve the strong non-linear merger stage of the collision using a
numerical relativity code solving Einstein's equations in the ADM
formalism as described in Ref.\cite{Baker:2002qf}. Our simulations are
limited to be accurate for less than $15M$ in time, where M is total
ADM mass of the system.  Despite these limitations we are able to
maintain a grid spacing of $M/24$ near the black holes.  Making use of
adapted ``fisheye'' coordinates, our spatial domain extends across
about 70M, so that we can avoid the delicate issue of modeling the
boundary by ignoring the regions of spacetime in the causal future of
the boundary.  At later time, after the black holes have merged into a
single distorted rotating black hole we apply the Lazarus approach to
describe the ring-down of the finally formed black hole. The linear
momentum is computed as in Ref. \cite{Campanelli99}.  The results for
$q=1.00, 0.90, 0.83, 0.77, 0.71, 0.66, 0.60$ are summarized in Fig.\
\ref{fig:PxM}. We added a point for $q=0.50$ computed with newly
developed improved full numerical techniques using a more stable
formulation of Einstein's equations \cite{Lousto:2004}. For
completeness we also show the estimates of the recoil 
velocity obtained by extrapolating perturbation theory results for the
plunge phase using the Fittchet's scaling function $f(q)$
as described in Ref.\cite{Favata:2004wz}.

It is worth to note here that the addition of the effects of spins of the
individual black holes may greatly affect the estimates derived here
for non-spinning black holes. Finally, more accurate numerical 
simulations, perhaps using higher order finite difference codes 
and/or adaptive mesh refinements techniques, are needed to
narrow down the current error bars and explore the $q<0.6$ region. 

%\subsection{Acknowledgments}
\ack
The results discussed in this paper have been obtained in
collaboration with John Baker, Carlos Lousto and Ryoji Takahashi. I
thank my collaborators John Baker, Carlos Lousto, 
and Marc Favata and David Merritt for very useful
comments to this manuscript. I also thank Harald Pfeiffer for making
available to us results from his Ph.D thesis. We gratefully
acknowledge the support of the NASA Center for Gravitational Wave
Astronomy at The University of Texas at Brownsville (NAG5-13396) and
also from NSF from grants PHY-0140326 and PHY-0354867.

\vskip 12pt
\providecommand{\bysame}{\leavevmode\hbox to3em{\hrulefill}\thinspace}
\providecommand{\MR}{\relax\ifhmode\unskip\space\fi MR }
\providecommand{\MRhref}[2]{%
  \href{http://www.ams.org/mathscinet-getitem?mr=#1}{#2}
}
\providecommand{\href}[2]{#2}

\end{document}